# Origin of charge density at $LaAlO_3$-on-$SrTiO_3$ hetero-interfaces; possibility of intrinsic doping

**Wolter Siemons[1,2,*], Gertjan Koster[1,*], Hideki Yamamoto[1,3], Walter A. Harrison[1], Gerald Lucovsky[4], Theodore H. Geballe[1], Dave H.A. Blank[2] and Malcolm R. Beasley[1]**

*[1] Geballe Laboratory for Advanced Materials, Stanford University, Stanford, California, 94305, United States of America. [2] Faculty of Science and Technology and MESA+ Institute for Nanotechnology, University of Twente, 7500 AE, Enschede, The Netherlands. [3] NTT Basic Research Laboratories, 3-1 Wakamiya Morinosato, Atsugi-shi, Kanagawa, 243-0198 Japan. [4] Physics Department, North Carolina State University, Raleigh, North Carolina, 27695, United States of America. [*]Contributed equally to this work*

*Correspondence should be addressed to gkoster@stanford.edu [G.K.]*

## Abstract

**As discovered by Ohtomo *et al.*, a large sheet charge density with high mobility exists at the interface between $SrTiO_3$ and $LaAlO_3$. Based on transport, spectroscopic and oxygen-annealing experiments, we conclude that extrinsic defects in the form of oxygen vacancies introduced by the pulsed laser deposition process used by all researchers to date to make these samples is the source of the large carrier densities. Annealing experiments show a limiting carrier density. We also present a model that** **explains the high mobility based on carrier redistribution due to an increased dielectric constant.**

Recently, Ohtomo and Hwang [1, 2] reported the existence of a conducting electron layer at the hetero-interface between two nominal insulators, $SrTiO_3$ and $LaAlO_3$. This is a remarkable result and has intrigued many researchers in the field [3, 4]. Equally remarkably, Ohtomo and Hwang [1, 2] found that for the $TiO_2$/LaO interface between $SrTiO_3$ and $LaAlO_3$ has a sheet carrier density of ~$10^{17}$ electrons/$cm^2$ and a mobility of $10^4$ $cm^2V^{-1}s^{-1}$, as inferred from conductivity and Hall-effect measurements; each of these is strikingly large.

Some insight into the possible sheet charge densities at a $SrTiO_3$/$LaAlO_3$ interface can be seen from the following considerations, which relate to an *intrinsic* doping mechanism. $SrTiO_3$ consists of charge neutral SrO and $TiO_2$ layers, whereas the $AlO_2^-$ and $LaO^+$ layers in $LaAlO_3$ have net charge and for an ideal planar interface yield a net interface charge equal to half that of the last plane [5]. Indeed a neutralizing charge at the interface is required to avoid a *polarization catastrophe* that arises due to this net interface charge. If left uncompensated, the energy associated with this polarization grows indefinitely as the thickness of the $LaAlO_3$ layer increases. Therefore, electrons have to be promoted to the conduction band of $LaAlO_3$ at some point. The charge that is necessary to prevent this polarization catastrophe is equal to half an electron per unit cell, one is led to 1/2 of an electron per unit cell or 3.2 x $10^{14}$ $cm^{-2}$. Note that this estimate applies only for perfectly stoichiometric $LaAlO_3$ and in this sense is an approximate upper limit in the intrinsic case. Any defects will reduce this number. In any event, clearly this line of reasoning cannot explain the very large charge densities observed. Of course, in the case of lower carrier densities, an *intrinsic* doping mechanism may become operative. Hence, an important subsidiary question is whether experimental conditions can be identified for which the intrinsic limit can be achieved.

The high sheet conductivities and carrier densities found by Ohtomo and Hwang have been linked in preliminary arguments by us and by others to possible oxygen vacancies (i.e., *extrinsic* doping) in the $SrTiO_3$ substrate. We further suggested that these vacancies are a result of high energy particle bombardment during the Pulsed Laser Deposition (PLD) process [6, 7] used so far by all workers in the field to make these hetero-structures. In addition, we proposed that the high mobility results from the thermal distribution of electrons away from the interface (consistent with Poisson's equation) where the intrinsic (dopant free) high mobility of $SrTiO_3$ is available. As we shall see, because of the very large dielectric constant of $SrTiO_3$ at low temperatures, this distribution can reach a large distance into the $SrTiO_3$.

In this Letter, we present a much more thorough study of the nature and origins of this conducting layer that puts our initial proposal on much firmer ground. We also show that annealing these interfaces at elevated temperatures in oxygen leads to a greatly reduced carrier density that is close in magnitude to that expected on simple ground from an ideal interface. More explicitly, transport, *in situ* UPS, Near Edge X-ray Absorption Spectroscopy (NEXAS) and visible to vacuum UV-Spectroscopic Ellipsometry (vis-VUV-SE) measurements have been performed on samples prepared under different oxidation conditions. From these experiments it is clear that, over a wide range of growth conditions, the large sheet charge density observed at these hetero-interfaces is due almost certainly to oxygen vacancies (donating electrons) in the $SrTiO_3$ substrate.

All of the films reported here were grown using Pulsed Laser Deposition (PLD) as described previously [6]. During growth, Reflection High Energy Electron Diffraction (RHEED) was used both to determine the amount of material deposited and to monitor the morphology of the samples. RHEED intensity oscillations reveal that typically 120 laser pulses are required to grow a monolayer of $LaAlO_3$. We found that if atomic oxygen is introduced during deposition, the growth proceeds in a multilayer fashion, as revealed by the damping of the RHEED oscillations, and thus could complicate the transport properties. Therefore, in order to achieve high degrees of oxidation without such degradation, atomic oxygen was only introduced after deposition during sample cooling, or after taking the sample out of the system for characterization and subsequently reintroducing it for a post oxidation treatment.

After deposition, the samples were moved *in situ* into an adjacent photoemission analysis chamber ($<5\times10^{-10}$ Torr base pressure) where their electronic structure was studied. Electrical transport properties were measured *ex situ* with a Quantum Design Physical Properties Measurement System (PPMS) using the Van der Pauw geometry, taking appropriate precautions to avoid photo-induced carriers. NEXAS and vis-VUV-SE are described elsewhere [[8, 9]].

In Table 1, we first summarize the results of transport measurements on two classes of samples. The first was prepared under relatively low oxidation conditions ($10^{-6}$ Torr, as measured with a hot cathode ion gauge), resulting in a high number of carriers [[10]]. The second was deposited and cooled at higher oxidation conditions ($2\times10^{-5}$ Torr), resulting in a reduced carrier density. Within each class, samples with $LaAlO_3$ thicknesses of 1 and 5ML were studied, the thinner samples being used for surface sensitive measurements. The transport data are very much in line with what has been reported by us and others, and demonstrate directly that the oxygen pressure during deposition clearly affects the transport properties of these hetero-structures [[11]].

As discussed previously in Ref [6], analysis of the UPS data on the 1ML $LaAlO_3$ samples shows the existence of a finite density of states at the Fermi energy that is sensitive to the oxidation history of the sample, being lowered by further oxidation. A measure of the density of states near the Fermi energy was determined by subtracting from the UPS data a background function given by the overall trend in the data from -3 eV to +2 eV. Integrating the density of states so obtained over energy to average the noise, one obtains a relative measure of the number of carriers per unit volume in the conduction band of the $SrTiO_3$. The results are shown in column 5 of Table 1 where the result for the sample fabricated at low oxidation conditions was set arbitrarily at 100%. The values found for various samples were correlated with their respective deposition oxidation conditions as expected from our arguments. A comparison of the density of states produced using various deposition conditions was reported elsewhere [6].

The NEXAS O $K_1$ edge absorption results are shown in Fig. 1(a). As is evident, the Ti 3d and 4s features in the sample prepared at high oxidation conditions are significantly stronger relative to the La 5d features. The half-width-at-half-maximum (hwhm), at the low energy side of the Ti $T_{2g}$ feature relative to $TiO_2$ is smaller for the sample prepared at high oxidation conditions compared with the sample at low oxidation conditions. The actual values are shown in Table 1. This is indicative of decrease in the density of O-atom vacancies, or alternatively $Ti^{3+}$ bonding. Due to the presence of oxygen defects all features in the NEXAS spectrum tend to broaden and therefore affecting all peaks in the spectrum. The absorption constants, $\alpha$, in Table 1, extracted from vis-VUV-SE measurements in Fig. 1(b) support the interpretation of the of the O $K_1$ spectra. The band edge defect state features between 2.3 and 3.5 eV at the onset on conduction band Ti 3d absorption edge, are stronger in the sample deposited at the lower oxygen pressure consistent with a higher concentration of O-atom vacancies, or equivalently $Ti^{3+}$ defect sites.

The data presented in Table 1 confirm the strong dependence of the properties of the conducting layer at these $SrTiO_3/LaAlO_3$ interfaces on the oxidation conditions under which the samples were made. More importantly, they correlate these results, through the bonding conditions of the Ti, to oxygen vacancies in the $SrTiO_3$.

We turn now to the question of how low the carrier density can be made at these interfaces. In order to address this question, we annealed various samples in atomic oxygen at elevated temperatures (350-800 °C). The results of these measurements are shown in Fig. 2. To exclude the possibility of introducing interstitial oxygen, we cooled the samples below 250 °C in vacuum, after annealing in vacuum for one hour at that temperature. The samples annealed up to 800 °C show a reduced sheet carrier density: 1.2-1.3 x $10^{13}$ down from 2.3 x $10^{13}$ at 4 K, whereas the mobility remains unchanged. Most strikingly, for anneals above 500 °C, the carrier sheet density appears to settle at a lower limit value and becomes independent of temperature, consistent with the findings of the Augsburg group [4]. Clearly these suggestive results warrant further investigation.

Regardless of the mechanism that created the carriers, there remains the question of what causes the high mobility, what causes the temperature dependence and where the carriers reside. The carriers cannot be distributed uniformly over the bulk unless there were an equal density of compensating charged donors, and then the mobility would be much lower than we observe [12]. The other extreme would be that they occupied surface bands at the interface, but there is no evidence to support this and we so no reason to suggest that such bands exist for this type of interface. The more plausible scenario is that this is a kind of modulation doping, where a high mobility conducting layer results because the carriers are separated in space from the charged vacancies due to their thermal distribution in space.

To support this scenario, we have modeled the distribution of the electrons in the $SrTiO_3$. We assume that the vacancies are distributed homogeneously in a narrow region of depth 6 nm in the $SrTiO_3$ at the interface. We assume a Thomas-Fermi distribution of the carriers, which is valid when $V(x)<kT$, appropriate for the cases that follow. Using this source distribution, we solve Poisson's equation to obtain the potential and charge density as a function of depth in the $SrTiO_3$, which is linked to the carrier density at a depth $x$ by $n(x)=-n_0V(x)^{3/2}$, with $n_0=(2m/\hbar^2)^{3/2}/3\pi^2$ and $V(x)$ being the potential at depth $x$. The outcomes for both a high ($10^{16}$) and the low ($10^{14}$) carrier density cases are shown in Fig. 3 at 300 and 4K.

The calculations show that the electrons move into the material over 50 nm at low temperatures and that this depth decreases dramatically at high temperatures (see dotted lines in Fig. 3) – a rather counterintuitive result. The reason for this unusual behavior is the very large and strongly temperature dependent dielectric constant of $SrTiO_3$, some 20,000 at low temperature compared to 300 at room temperature [[13],[14]]. It also offers an explanation of the high mobility that is measured especially for the samples deposited at low pressures: the carriers are moved away from the defects into the pristine $SrTiO_3$, where they are highly mobile. Furthermore, it explains the absence of Shubnikov-de Haas oscillations, since the carrier density varies with depth.

In summary, measurement of the electronic properties of the interfaces created by depositing $LaAlO_3$ on $SrTiO_3$ show electronic properties similar to the remarkable values found originally by Ohtomo and Hwang [[1]]. Also, UPS spectra show states at the Fermi level, indicating a conducting interface. The number of these states is lowered when the sample is oxidized, suggesting that oxygen vacancies play an essential role in supplying the charge carriers. This is further confirmed by NEXAS and vis-VUV-SE measurements, which show more $Ti^{3+}$ for samples made at lower pressures. We argue that the vacancies are created by the PLD process itself where relatively high energy particles sputter off oxygen. To reduce the number of vacancies we have annealed samples in atomic oxygen, which reduces the number of carriers, but keeps the mobility the same. The dependence of sheet carrier density as a function of temperature is changed dramatically. To determine where the electrons are located we have calculated the potential and the carrier density in the $SrTiO_3$ as a function of distance from the interface and calculated that the electrons move into the pristine $SrTiO_3$ over large distances, mainly due to the high dielectric constant of $SrTiO_3$ at low temperatures.

**Acknowledgements**

One of us (G.K.) thanks the Netherlands Organization for Scientific Research (NWO, VENI). W.S. thanks the Nanotechnology network in the Netherlands, NanoNed. We also wish to acknowledge M. Huijben, T. Claeson, R.H. Hammond, J. Mannhart, H. Seo, J. Lüning and A.J.H.M. Rijnders. This work was supported by the DoE BES with additional support from EPRI.

TABLE 1. A comparison between samples made at low and high pressures. The most important parameters, measured with different techniques, are shown.

FIG 1. a) NEXAS $OK_1$ of sample 0728 (deposited at $10^{-5}$ Torr) and sample 0802 (deposited at $10^{-6}$ Torr) b) vis-VUV Spectroscopic Ellipsometry absorption constant, $\alpha$, spectrum of the same samples as in a).

FIG 2. Sheet carrier densities at 20 K (blue symbols) and 300 K (red symbols) as a function of annealing temperature in 600 W atomic oxygen, for samples made at $10^{-5}$ Torr of $O_2$ 600 W of atomic oxygen corresponds to ~$10^{17}$ oxygen atoms $cm^{-2}$ $s^{-1}$. The latter value was taken after Ingle et al. [[15]] who worked on the same system in our laboratory. The values at 25 °C indicate the as deposited samples. The different symbol shapes indicate different samples made under similar conditions: two made at Stanford (circles and triangles), the other two made at the University of Twente (crosses and squares). Sample thickness ranges from 5 to 26 ML.

FIG 3. Calculations of the electron density as a function of distance from the interface for two different total numbers of electrons per $cm^2$ ($3\times10^{14}$ and $1\times10^{16}$) at two different temperatures. The shaded blue area indicates the location of the oxygen vacancies for the high carrier density case, $10^{16}$, for the low carrier density case no oxygen vacancies were assumed, hence the different slopes at the interface.

Table 1 Siemons et al.

| Prep. cond. ($pO_2$ in Torr) | Thickness [ML $LaAlO_3$] | $n_s$ [$cm^{-2}$] 4K/300K | $\mu$ [$cm^2 V^{-1}s^{-1}$] 4K/300K | UPS (relative int. at $E_F$) | NEXAS $\delta$(hwhm) $Ti_{2g}$ ** | Vis-VUV-SE $\Delta(3.5\text{-}2.5)^{***}$ |
|---|---|---|---|---|---|---|
| $10^{-6}$ | 1 | N/A | N/A | 100% | N/A | N/A |
| $10^{-6}$ + 6000L* | 1 | $2\times10^{16}$ / $2\times10^{16}$ | $1\times10^{4}$ / 5 | 40% | 1.3 | 3.9 |
| $10^{-6}$ | 5 | $2\times10^{16}$ / $2\times10^{16}$ | $1\times10^{4}$ / 5 | N/A | N/A | N/A |
| $10^{-5}$ | 1 | Not conducting | Not conducting | N/A | 1.1 | 5.8 |
| $10^{-5}$ | 5 | $2\times10^{13}$ / $2\times10^{14}$ | $3\times10^{2}$ / 4 | N/A | N/A | N/A |

* 1 ML sample prepared at $10^{-6}$ Torr of $O_2$ and after *in situ* exposure to $10^{-6}$ Torr of $O_2$ for 10 min. at 150 °C

** ratios of hwhm of $T_{2g}$ d-state relative to $TiO_2$

*** ratios of $\alpha$ between 3.5 and 2.5 eV

Figure 1. W. Siemons *et al.*

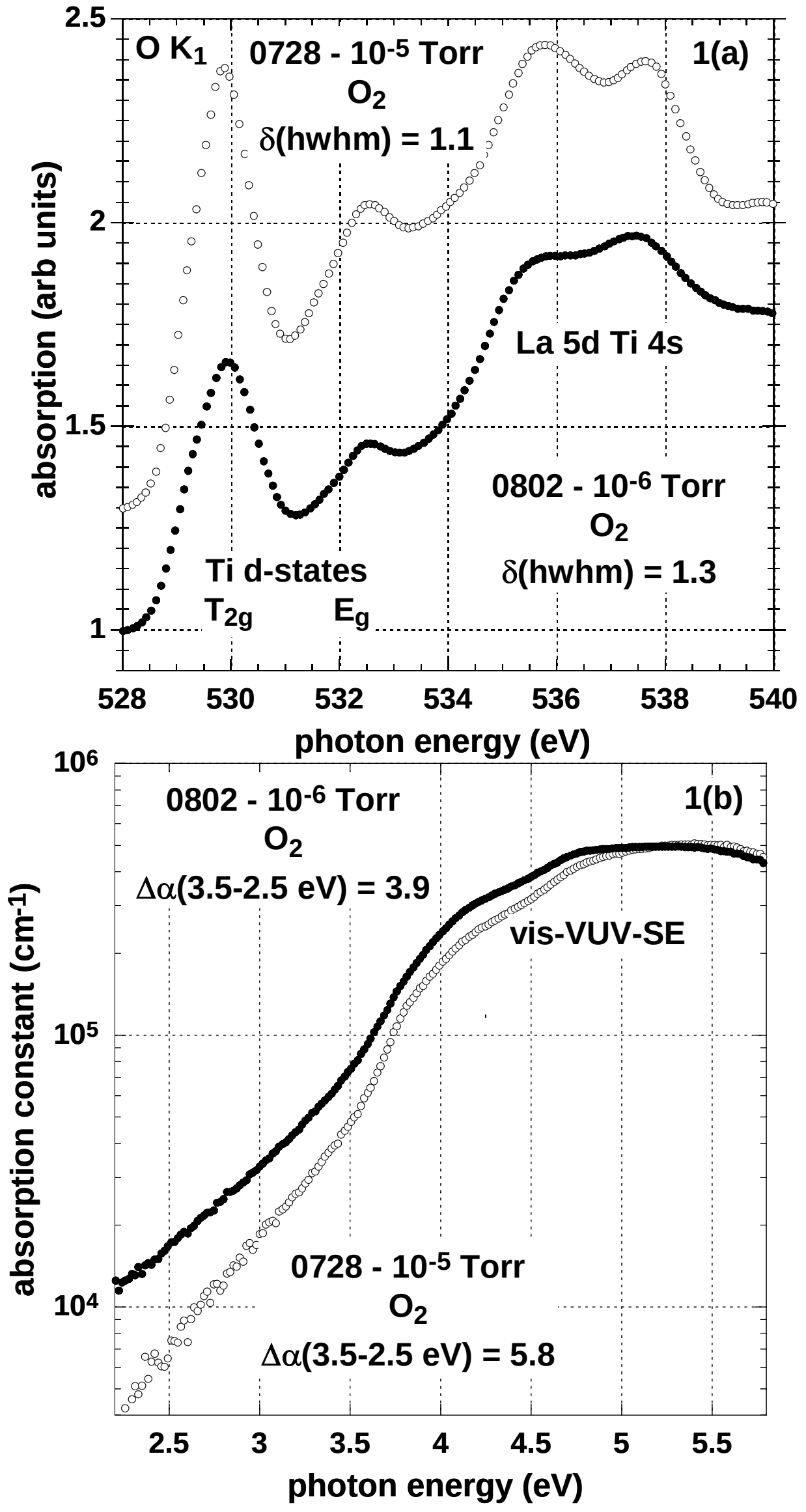

Figure 2. W. Siemons *et al.*

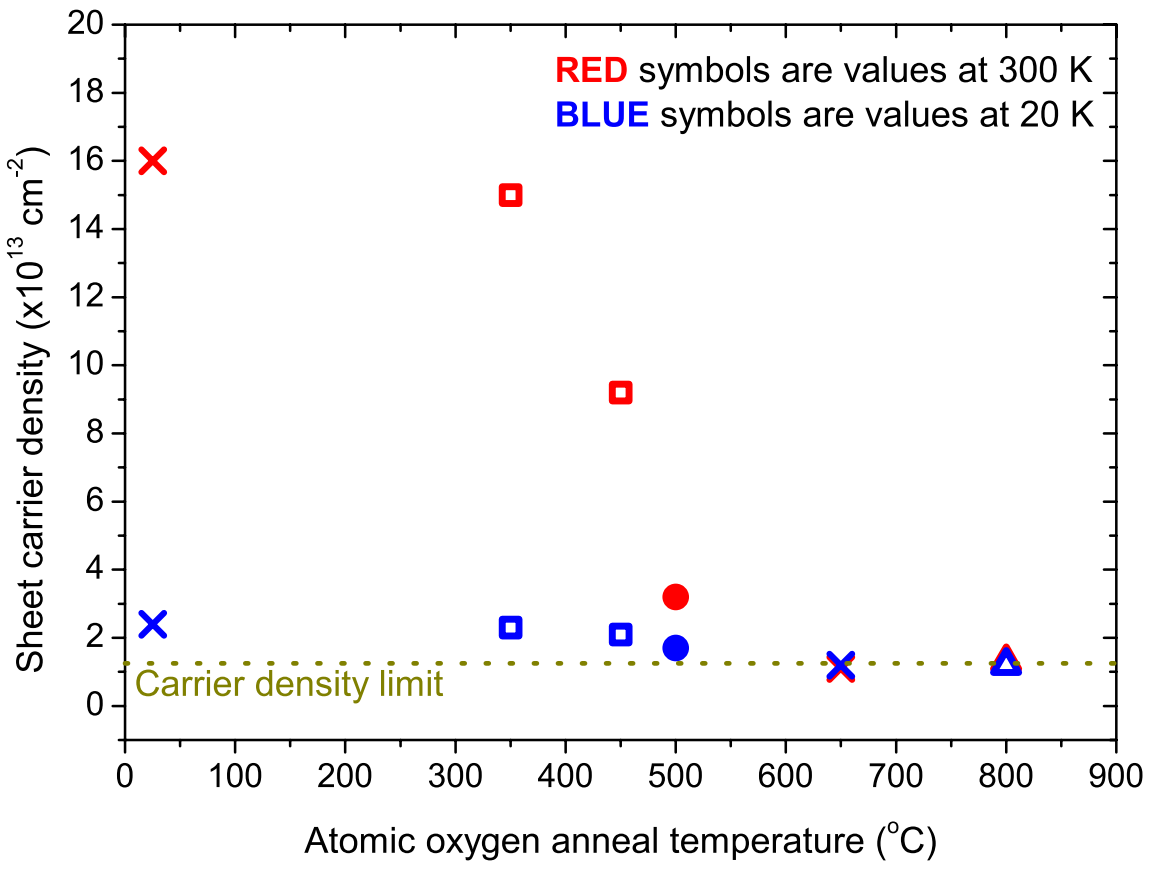

Figure 3. W. Siemons *et al.*

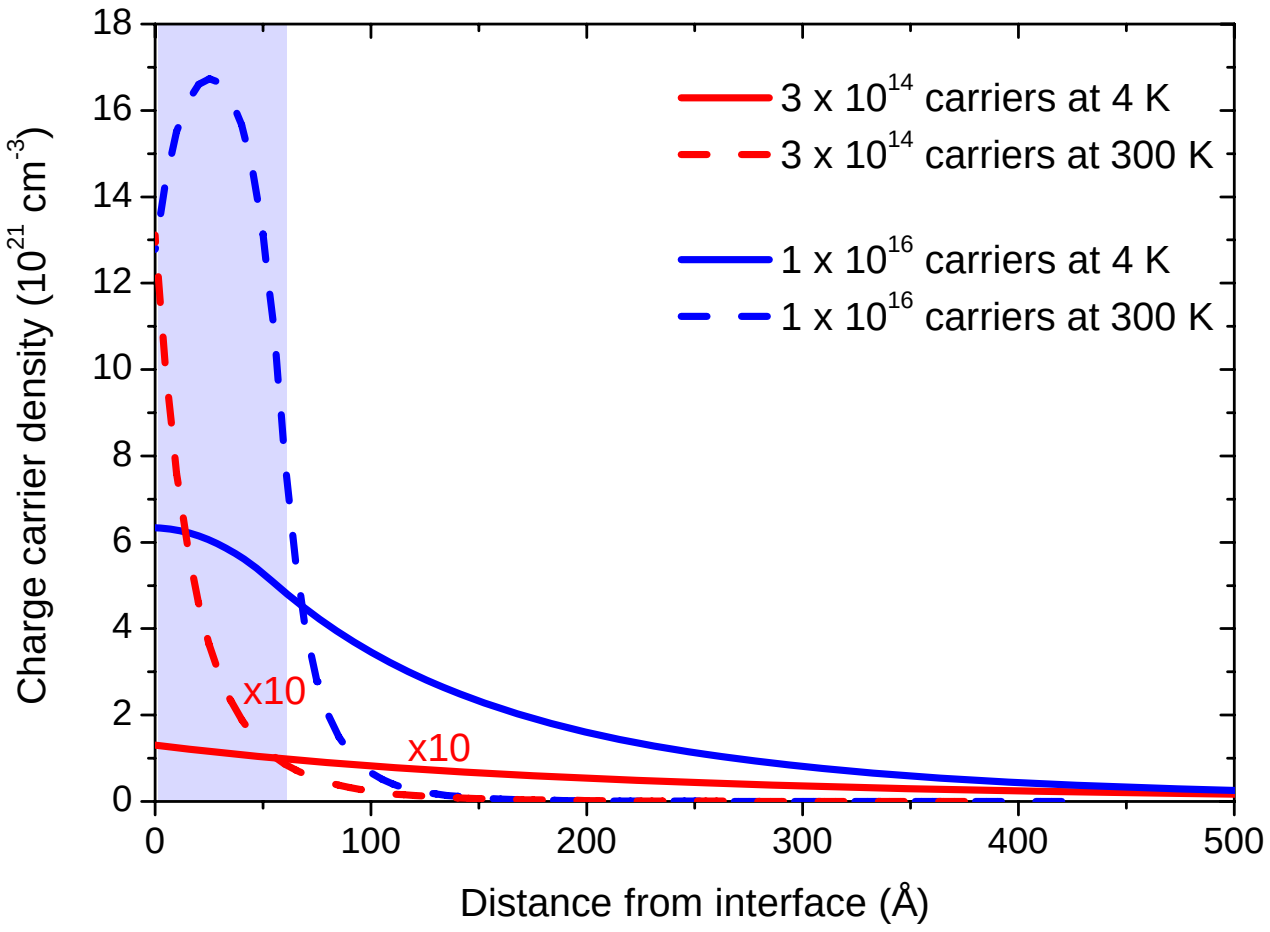